%% file: paper.tex
\documentclass[a4paper,fleqn]{cas-dc}

\input{preamble}
\begin{document}
\let\WriteBookmarks\relax
\def\floatpagepagefraction{1}
\def\textpagefraction{.001}

\shorttitle{Assessment of the N-1 voltage security of a future Nordic energy system}

\shortauthors{Kuhrmann et~al.}

\title[mode = title]{Assessment of the N-1 voltage security of a future Nordic energy system}

\author[1]{Luis Kuhrmann}[type=editor,
                        auid=000,bioid=1,
                        orcid=0000-0001-7143-4576]

\cormark[1]

\ead{kuhrmann@chalmers.se}

\credit{Investigation, Methodology, Software, Writing - Original draft, Visualization}

\affiliation[1]{organization={Division of Energy Technology, Chalmers University of Technology},
    city={Göteborg},
    country={Sweden}}

\author[2]{Peiyuan Chen}
\credit{Conceptualization, Supervision, Methodology, Writing - Review \& Editing}

\author[1]{Lisa Göransson}
\credit{Conceptualization, Supervision, Writing - Review \& Editing}

\affiliation[2]{organization={Division of Electric Power Engineering, Chalmers University of Technology},
    city={Göteborg},
    country={Sweden}}

\author%
[1]
{Filip Johnsson}
\credit{Supervision, Writing - Review \& Editing, Funding acquisition}

\cortext[cor1]{Corresponding author}

\include{Abstract}


\begin{NoHyper}
\maketitle
\end{NoHyper}

\input{Introduction}

\input{Method}

\input{GridModel}

\input{Results}

\input{Discussion}
\input{Conclusion}

\section*{AI declaration}
During the preparation of this work the authors used ChatGPT (OpenAI) in order to improve the language and sentence formulation of the manuscript. 
The authors subsequently reviewed and edited the content and take full responsibility for the content of the publications.
No AI tools were used in the development of the associated software.

\input{Appendix}

\section*{Acknowledgements}
\noindent
North European Energy Perspectives Project, P2023-00759.
\printcredits

\bibliographystyle{model1-num-names}
\bibliography{smaller_bibfile}

\end{document}

%% file: preamble.tex
\usepackage{circuitikz}
	
\ctikzset{sources/scale=2}
\ctikzset{grounds/scale=1.5, grounds/thickness=1.5}

\usepackage{subcaption} %
\usepackage[capitalize]{cleveref} %

\AddToHook{cmd/appendix/before}{\crefalias{section}{appendix}}
\AddToHook{cmd/appendix/before}{\crefalias{subsection}{appendix}}

\usepackage{placeins} %

\usepackage[numbers]{natbib}

\usepackage[inkscapelatex=false]{svg} %
\usepackage{xurl} %

\newcolumntype{Q}[1]{>{\centering\let\newline\\\arraybackslash\hspace{0pt}}m{#1}}

%% file: Abstract.tex
\begin{abstract}
Capacity Expansion - Energy System Models (CE-ESMs) have been widely used to optimize the decarbonization of future energy systems on national and continental scales. 
To capture the limitations in electricity trade between different parts of the system investigated, CE-ESMs often include a representation of the transmission grid.
However, CE-ESMs usually only include linearized representations of the grid and omit or strongly simplify grid stability requirements.
In this work, we evaluate the N-1 voltage security of the electricity system results obtained from a CE-ESM of a decarbonized Nordic energy system in Year 2050, and analyze the effects of different grid code requirements and shunt capacitor and reactor automation on voltage security. 
For comparison, we evaluate a Year 2022 system based on ENTSO-E data.
We show that the CE-ESM results for the Nordic countries are not N\nobreakdash-1 voltage-secure and that, depending on the grid code and shunt automation requirements, the contingencies with voltage violations range from 1.1\% to 17.4\%. 
Furthermore, the results indicate that both Power Park Module voltage control and shunt extreme voltage automation have strong positive effects on voltage security in the modeled future system.
Grid regions with low levels generation, and thus little dynamic reactive power from generators, are found to be the most at risk of being voltage-insecure.

\end{abstract}

%% file: Introduction.tex
\section{Introduction}

Decarbonization of energy systems is a necessary measure to tackle climate change. 
The electrification of various sectors, including industry and transportation, is central to decarbonization. 
As a consequence, the demand for electricity is expected to increase in the future. 
Reductions in the costs for renewable electricity sources, such as wind and solar power, allow the cost-effective supply of these electricity demands~\cite{IRENA_2026}.
Capacity Expansion~- Energy System Models (CE-ESMs) have been widely used to study different future electrification scenarios in academia, policymaking, and industry~\cite{Motta_2024,TYNDP_2024}.
Modern CE-ESMs capture both investment and dispatch, allowing optimal management of the variations inherent to wind- and solar-powered electricity production~\cite{Motta_2024}.
CE-ESMs primarily use mathematical optimization to find the cost-optimal capacities of various electricity generation and storage technologies, in order to meet a given demand for electricity, as well as, in many instances, the given demands for other energy carriers, such as heat and hydrogen.

The electricity grid is an important element of CE-ESMs, as the transmission of electricity allows for the balancing of power variations over large geographic areas and for attaining the optimal combination of different renewable energy resources. 
CE-ESMs have traditionally entailed strongly simplified representations of the electricity grid, only including major bottlenecks between national borders or bidding zones~\cite{Brown_2018a,TYNDP_2024,Goeransson_2023a}.
Despite the difficulties associated with high-resolution CE-ESM modeling~\cite{Javanmardi_2025}, recently the spatial resolution of some CE-ESMs has increased to the point that individual transmission lines can be modeled, which allows for more accurate consideration of local bottlenecks~\cite{Lindner_2025,Frysztacki_2021,Bertilsson_2026}.

Grid stability is a critical requirement for grid structure and operation.
The stability aspects of grids with high shares of inverter-connected generation (i.e., wind and solar generation) can be grouped into the following categories:
frequency stability; rotor angle stability; voltage stability; resonance stability; and converter-driven stability~\cite{Hatziargyriou_2021}. In addition, the thermal capacity limitations of transmission lines and other equipment must be respected.
Of these aspects, converter-driven stability is largely affected by the control design of the converter-interfaced Power Park Modules (PPMs) and their interactions with a weak grid connection. However, at the capacity expansion planning stage, the specific converter design is usually unknown, which makes it difficult to conduct such a study at the planning stage. 
While resonance stability is influenced by the system configuration, it is primarily dependent upon generation unit design and control.
Both converter-driven stability and resonance stability are topics for future research. 
On the other hand, frequency stability is a global phenomenon that is dependent upon the system composition and available ancillary frequency services, with little dependence on the placement of generation technologies within the system. 
Therefore, frequency stability is not as critical to consider when trying to find the optimal spatial system configuration. 
However, voltage stability, rotor angle stability, and thermal limits are heavily dependent upon the spatial system configuration and operation, with the placement and dispatch of generation units being crucial factors.
Thus, voltage stability, rotor angle stability, and thermal limits should be considered when trying to find an optimal spatial system configuration. %
To limit the scope of this work, we focus exclusively on voltage stability. %

To contain the voltage of the transmission grid within a narrow band around its nominal voltage at all times, static and dynamic reactive power sources are used. 
Static reactive power sources, primarily consisting of shunt capacitors and shunt reactors, are operated in steps; they do not allow for continuous operation. 
While they are generally slower, they are the cheapest source of reactive power and are generally used to provide bulk supplies of reactive power. 
In contrast, dynamic reactive power sources, such as generators or Flexible AC Transmission System (FACTS) devices, are continuously regulable in their reactive power output and generally respond faster than shunt capacitors or reactors. However, sources of dynamic reactive power are generally more expensive than sources of static reactive power. 
In general, sources of static reactive power are used to coarse-tune the voltage and to adjust for slower changes in flow patterns, while sources of dynamic reactive power are used to fine-tune the voltage and to respond quickly to rapidly occurring changes, such as contingencies.

Some aspects of voltage limitations have been captured in CE-ESMs~\cite{Allard_2020,Prianto_2026,Wogrin_2020}, where the need for reactive power to manage voltage variations is included.
However, these previous studies did not make a distinction between dynamic reactive power (from generators and FACTS devices) and static reactive power (from shunt capacitors and shunt reactors). 
Instead, dynamic reactive power and static reactive power (when included) are both used to fulfill the reactive power needs of the steady-state operational conditions. 
Furthermore, the reactive power needs in the case of a contingency are not considered.
In reality, the use of dynamic reactive power for static voltage balancing should be limited to ensure the availability of sufficient dynamic reactive power in the case of a contingency. 

It is crucial that the voltage of the transmission grid is able to recover quickly to its nominal range after any individual contingency, so as to avoid further voltage collapse; this is called \emph{N\nobreakdash-1 voltage security}.
A recent case in which the voltage was not sufficiently contained after changes in system configuration, which ultimately led to a blackout, is the 2025 Iberian Peninsula blackout \cite{entsoe_2026}.
One of the root causes of that blackout was the lack of dynamic reactive power during the contingencies.
The need for dynamic reactive power for contingencies has been widely modeled in security-constrained optimal power flow~\cite{Nycander_2019}.
However, this has in general been achieved in a non-linear fashion and only as part of operational dispatch models.
During contingencies, dynamic reactive power needs to be available in electrical proximity to the locations with fluctuating voltages.
The dynamic reactive power is supplied primarily by generation units. %
However, if the reactive power supplied by the generation units is electrically too distant, investments in expensive FACTS devices may be needed to provide sufficient local dynamic reactive power.
Therefore, the dynamic reactive power needs for contingencies should be considered when identifying the cost-optimal system configuration.

An additional factor in the available dynamic reactive power from generation units, beyond their location, is how much and when they supply dynamic reactive power.
Minimum requirements for available dynamic reactive power are placed on generation units using grid codes, typically in terms of P-Q capability diagrams and V-Q profiles.
Currently, many grid codes do not require PPMs, e.g., wind and solar power generation units, to be in voltage control mode, which is a prerequisite for providing dynamic reactive power in the case of a contingency~\cite{EIFS_2018,Energinet_2025}.  
Additionally, PPMs are generally not expected to provide dynamic reactive power when not producing active power, and requirements as to their voltage control point positions are usually project-specific.
However, as the share of PPMs is expected to increase, the requirements of current grid codes may be insufficient.
Thus, various changes in grid codes have been suggested or implemented, including the requirement of voltage control as a default setting, a change of control point, and/or the requirement to provide dynamic reactive power even when not producing active power~\cite{SvK_utkast_2024,Stattnet_2024}.  

Other units that provide reactive power support after an N-1 contingency event are voltage-controlled switched shunts. Historically, the switching of shunt capacitors or reactors has been manually controlled, which is still the case in some transmission systems~\cite{entsoe_2026}. However, many shunt capacitors and reactors are now equipped with voltage automation, allowing for automatic switching in cases of extreme voltages. For example, extreme-voltage automation (EXA) is implemented in the Swedish transmission system~\cite{Svk_2019}.
The fast reaction of static reactive power sources can have a significant impact on voltage stability, and its non-existence in the Spanish grid has been cited as an important factor in the ENTSO-E report on the 2025 Iberian Peninsula blackout~\cite{entsoe_2026}.

In this work, we model N\nobreakdash-1 post-contingency static voltage security for a future electricity system as given by a CE-ESM, using the Nordic synchronous transmission system (220--400~kV) in Year 2050 as a case study while considering the impacts of different grid codes and EXA configurations. We compare the results to the analysis of a historic scenario (Year 2022). 
The impacts of contingencies of AC and DC lines, generators, and loads on the voltage of the transmission grid are evaluated.

Our aim is to answer the following questions:
\begin{itemize}
\item Are the future systems created by CE-ESMs N\nobreakdash-1 voltage-secure?
\item If not, which type of contingencies are voltage-insecure in a future Nordic system as created by a CE-ESM? 
\item What influences do the reactive power requirements in PPM grid codes and extreme-voltage automation (EXA) systems exert on N\nobreakdash-1 voltage security? 
\end{itemize}

The main novelties and contributions of this work include:
\begin{itemize}
    \item We assess the N-1 static voltage security of challenging operating states obtained from a CE-ESM of a highly decarbonized Nordic energy system in 2050, for testing the feasibility of cost-optimized future system configurations against voltage security requirements that are typically omitted from capacity-expansion optimization.
    \item We quantify how the voltage security of a future system is affected by reactive power provision that are typically not modeled in CE-ESMs, including PPM voltage and reactive power control mode, alternative control points, different reactive capability requirements at the connection point, and extreme-voltage automation of shunt capacitors and reactors. 
    These assumptions change the fraction of voltage-insecure contingency under the analyzed flow states from 17.4\% to 1.1\%.
    \item We identify limited local dynamic reactive power capability as the root cause for the observed voltage-insecure problems in regions where the CE-ESM places little generation. 
    This indicates a specific link between long-term generation expansion decisions and N-1 post-contingency voltage security that is not captured by the CE-ESM.
\end{itemize}

%% file: Method.tex
\section{N-1 Contingency analysis} \label{sec:n-1_cont_analysis}

Overall, two key assumptions are made in the contingency analysis, in that before the contingency: 1) all reactive power is assumed to be provided by shunt capacitors and reactors; and 2) the voltages in the transmission grid are at 1 per unit (p.u.).

\subsection{Contingencies}\label{sec:contingencies}
We include four types of contingencies: HVAC line contingencies; HVDC line contingencies; generator contingencies; and load contingencies.

\subsubsection*{HVAC contingencies}
The tripping of each transmission line is included as a contingency. 
For double-circuit lines, we only consider the tripping of one of the circuits. 
We do not consider the loss of entire substations, as buses are typically configured to have redundant busbars. 
Transformers in the substation are also typically configured to be redundant and, thus, are not included as contingencies.

\subsubsection*{HVDC contingencies}
The tripping of each HVDC line in the system is included as a contingency. We assume that power transfer is lost and that voltage control and reactive power injection on both sides cease.

\subsubsection*{Generator contingencies}
Of the generation technologies, we only include nuclear power plants and offshore wind parks as contingencies.
Nuclear power plant contingencies are considered, as they are comprised of very large generation units the result in large flow changes in the case of a loss.
We consider a loss of active power generation of the size of the largest reactor in each nuclear power plant (0.51--1.45~GW), as well as the associated loss in reactive power support capability. 
Offshore wind park contingencies are considered, as they are often connected through a single connection, which can lead to the whole park tripping simultaneously, leading to large flow changes. We consider a loss of up to 1~GW of active power and the loss of the associated reactive power support capability under the assumption that offshore wind parks with capacities larger than 1~GW have several connection points.
We do not include other generation technologies as contingencies, as they are either smaller in size or have several connection points, so they have reduced contingency impacts.

\subsubsection*{Load contingencies}
As we do not have a breakdown of current electricity usage by industry with any spatial or temporal resolution, we cannot include the tripping of current large industrial loads, such as aluminum smelter halls, as contingencies. 
While this is unfortunate, current aluminum smelter halls are located close to HVDC lines, which cause larger disturbances when tripped, which should capture local grid limitations.

For the future scenario, our input data contain detailed information about the locations of new loads connected to the system. Of these new loads, only electrolyzers are sufficiently large and expected to be electrically concentrated, such that their tripping might result in instability. This stability-related risk from electrolyzers has also been expressed by the Nordic Transmission System Operators (TSOs)~\cite{NGDP_2025}. 

As the connection structure of future electrolyzers is currently unknown, we assume a maximum contingency size of 1~GW.

\subsection{Pre-contingency flow state selection} \label{sec:flow_state_selection}
To reduce the number of power flow analyses required to assess N-1 stability, only a subset of timesteps is evaluated.
Specific grid flow states are chosen as a consequence of extreme conditions, as these conditions are most likely to raise the risk of instability.
We choose 18 flow states by selecting the timesteps with the annual total minima and maxima for: \newline
generation; load; wind generation; nuclear power generation; hydropower generation; import/export; synchronous generation; net load; and grid transport.
In this context, the terms `import' and `export' refer to the total import/export of the Nordic synchronous system. We define net load as the difference between load and non-dispatchable generation (wind, solar, and nuclear power generation). We define grid transport as the power flow through a line multiplied by the length of the lines, summed over all the lines, which is also called the \emph{MW$\cdot$km measure}.

\subsection{Post-contingency AC power flow with FCR allocation} \label{sec:post_contingency_pf_and_FCR}
Individual contingency events are applied to the pre-contingency situation.
Next, Frequency Containment Reserve (FCR) is used to adjust for changes in generation or load, as described below. 
Then, the post-contingency power flow is run.

To improve post-contingency power flow convergence, we use two power flow tools: PowerModels.jl~\cite{Coffrin_2018}, which has been expanded to consider generation reactive power limits; and pandapower~\cite{Thurner_2018}. 
If either of them converges, we use that power flow result. 
If neither tool converges, we consider the simulation to be unstable and, thus, also voltage-insecure.

To model the power balance during the temporary steady state formed after the contingency and the frequency containment process, FCR is allocated to balance any loss of generation or load.
FCR is assumed to be provided exclusively by hydropower. 
All hydropower dispatch is uniformly and proportionally scaled either up or down to model the FCR response.

We make this assumption because 80\% of FCR in the Nordic countries is procured by Norway and Sweden~\cite{Entsoe_FCR_2024}.
In Sweden, around 60\% of capacity pre-qualified to provide FCR is hydropower~\cite{Svk_FCR_2026}, while in Norway, FCR-D has historically been procured through TSO decisions, thereby forcing hydropower to deliver FCR-D~\cite{Statnett_2023}.
In additionally, Nordic TSOs may procure up to one-third of their required FCR through trade with other TSOs.~\cite{Entsoe_FCR_2024}
As Norway and Sweden have significant hydropower resources that are well suited to FCR, it is not unlikely that Finland and Denmark procure some of their FCR from Norway and Sweden.
However, it should be noted that batteries are an increasingly important source of FCR provision, the influence of which on voltage stability is a topic for future work. %

\subsection{Evaluation of the results} \label{sec:evaluation_method}
We evaluate the post-contingency situation for voltage deviations. 
We follow the Swedish voltage limitations, as mentioned in~\cite{Svk_rapport_2024}. %
These voltage limitations consist of a recommended voltage range of  1.0--1.0375~p.u., a permanently admissible voltage range of 0.9875--1.05~p.u., and a temporarily admissible voltage range of 0.95--1.1~p.u.

As the pre-contingency situation in our model has unity voltages, which is not the case in reality, we derive more-conservative post-contingency voltage ranges from the Swedish voltage limitations.
The upper limit is set at 1.0625~p.u., which is computed by adding the margin between the upper temporarily admissible voltage limit (1.1~p.u.) and the upper recommended voltage (1.0375~p.u.) to the unity pre-contingency voltage. Similarly, the lower limit is set to 0.95~p.u., which is computed by subtracting the margin between the lower recommended voltage (1.0~p.u.) and the lower temporarily admissible voltage limit (0.95~p.u.) from unity.
When a transmission bus voltage is outside this range of 0.95--1.0625~p.u. following a contingency event, we consider it to be voltage-insecure.

\section{Modeling of reactive power support} \label{sec:modelling_reactive_q}
\subsection{Steady-state reactive power support} %
We assume that, in each flow state, slow-acting but cheap shunt capacitors and reactors are the primary means of regulating the voltage.
We make the assumption that this static reactive power support is ideally distributed and adjusted to achieve pre-contingency voltages of 1 p.u. at all transmission buses.
This is an optimistic and idealized assumption. 
In reality, the voltages at different buses are not uniformly controlled to their nominal value in the pre-contingency state. 
Nevertheless, the assumption is acceptable as our focus is on assessing the sufficiency of generators as dynamic reactive power sources during an N-1 contingency event.

Details and results regarding the needs for static reactive power compensation for the pre-contingency flow state can be found in \citet{Kuhrmann_2025}.

\subsubsection{Extreme voltage automation (EXA)}
Shunt capacitors and reactors in the Swedish transmission grid are equipped with EXA~\cite{Svk_2019}, i.e., they activate automatically once the voltage has been outside predetermined bounds for a predetermined (short) time. 
It is unclear as to whether similar systems exist in the rest of the Nordic system. 
For the purposes of this N-1 contingency analysis, we assume that the entire system either does or does not have EXA.

A \emph{No EXA} and \emph{With EXA} configuration is modeled. 
In the \emph{With EXA} configuration, the shunt capacitors and reactors available for activation are limited to installed shunt capacitors and reactors that are unused in the pre-contingency flow state. 
The installed shunt capacitors and reactors are determined by finding the required shunt capacitors and reactors for each pre-contingency flow state at each bus and then taking the annual maxima per bus to determine the required installation size per bus.

\begin{figure}
	\centering
		\includegraphics[width=\linewidth]{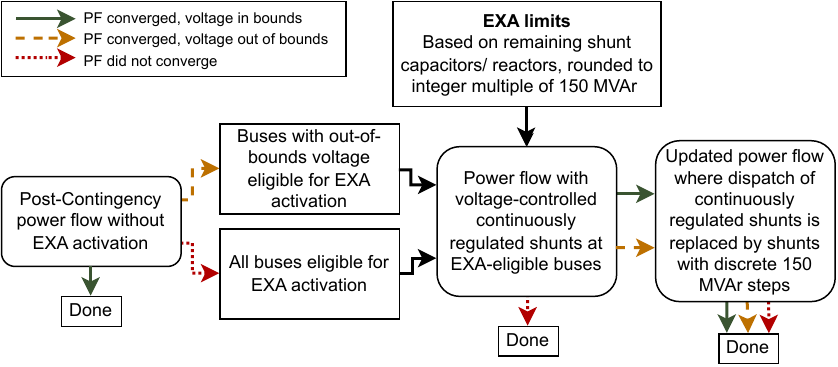}
	\caption{Flowchart of EXA activation.}
	\label{fig:flowchart_EXA}
\end{figure}

When modeling EXA, we consider shunt activation to be slower than the speed with which generators change their reactive power injections. 
First, generators change their reactive power injection. 
Then, if the transmission grid bus voltages remain outside of the voltage bounds, we consider EXA activation.
A typical unit size of 150~MVAr is assumed for EXA. 
A flowchart of the implementation is presented in \cref{fig:flowchart_EXA}.

\subsection{Availability of dynamic reactive power compensation}
We assume that only generators provide dynamic reactive power compensation. 
This neglects FACTS devices such as Static Synchronous Compensators (STATCOMs), which are already present in the Nordic transmission grid~\cite{SVK_2023}. 
While public information on the size and location of FACTS devices is severely limited, the amount of dynamic reactive power provided by currently installed FACTS devices is likely significantly lower than that provided by generation units. 

To model the available dynamic reactive power of the generators, we use the current minimum requirements of dynamic reactive power compensation capacity as specified by the grid codes in the specific Nordic country in which the generator is placed. We assume that all dynamic reactive power is available for the post-contingency situation. In reality, some of the dynamic reactive power is already used in the pre-contingency state.
For HVDC lines, we assume that Voltage-Source Converter (VSC) systems provide dynamic reactive power, as they can freely control their reactive power output, while we assume that Line-Commutated Converter (LCC) systems do not provide dynamic reactive power, as their reactive power output is not dynamically controllable.

\subsubsection*{Reactive power requirements on generators}  \label{sec:reactive_power_req}
\begin{figure}
\begin{subfigure}{\columnwidth}
	\centering
		\includegraphics[width=\columnwidth]{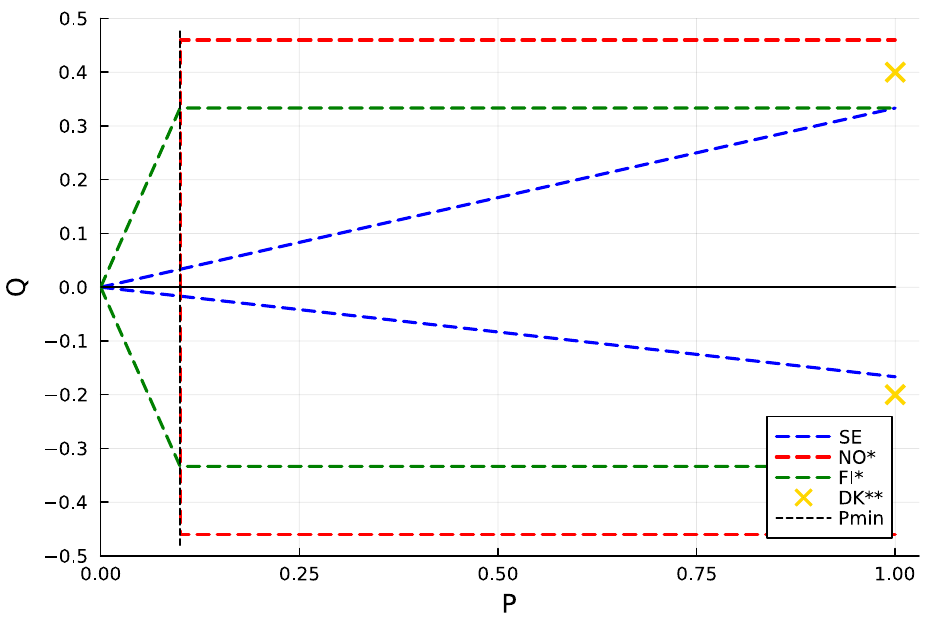}
	\caption{For synchronous generators}
	\label{fig:grid_code_sg}
\end{subfigure}
\begin{subfigure}{\columnwidth}
	\centering
		\includegraphics[width=\columnwidth]{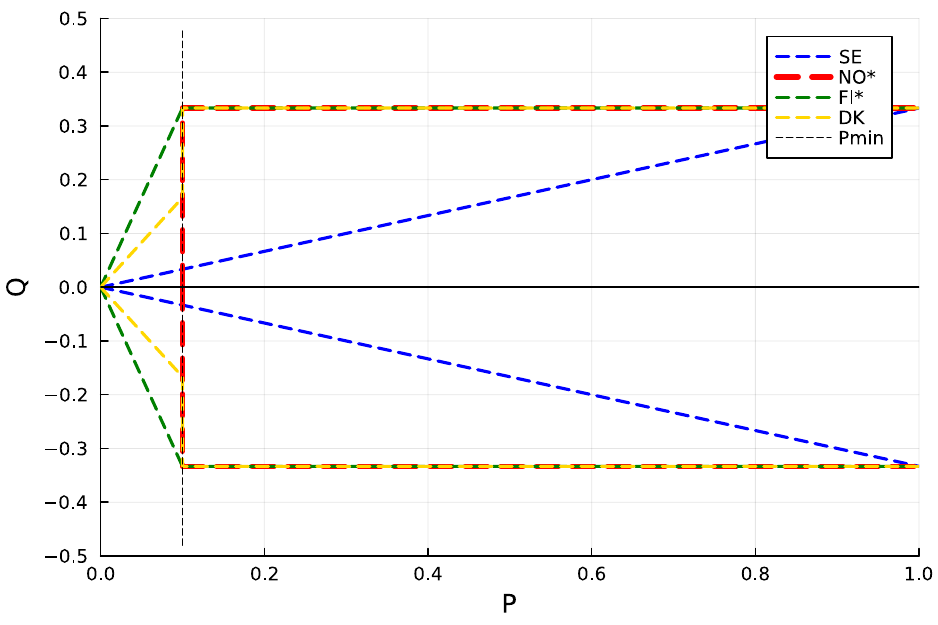}
	\caption{For power park modules}
	\label{fig:grid_code_ppm}
\end{subfigure}
\caption{Current minimum reactive power requirements as dependent upon the active power in the Nordic system~\cite{Stattnet_2024,Fingrid_2024,Energinet_2025,EIFS_2018}. *Pmin depends on the power plant. **Not defined for Energinet (DK).}\label{fig:grid_code}
\end{figure}

We assume that all generation units are of types C and D as defined by EU regulation, which in the Nordic synchronous area includes all generators with a maximum capacity of more than 10~MW~\cite{EU_2016}.
\Cref{fig:grid_code} summarizes the currently enacted requirements for minimum reactive power capability as a function of the active power output of synchronous generators (SG) and PPMs of types C and D.
PPMs refer to generators that are non-synchronously connected to the grid (i.e., inverter-connected generation).
The Swedish grid code differs from the other Nordic grid codes in terms of both SGs and PPMs by the reactive power requirement being proportional to the active power output ($P_\textrm{gen}$) rather than the maximum power output ($P_\textrm{max}$).

\begin{figure}
	\centering
		\includegraphics[width=\linewidth]{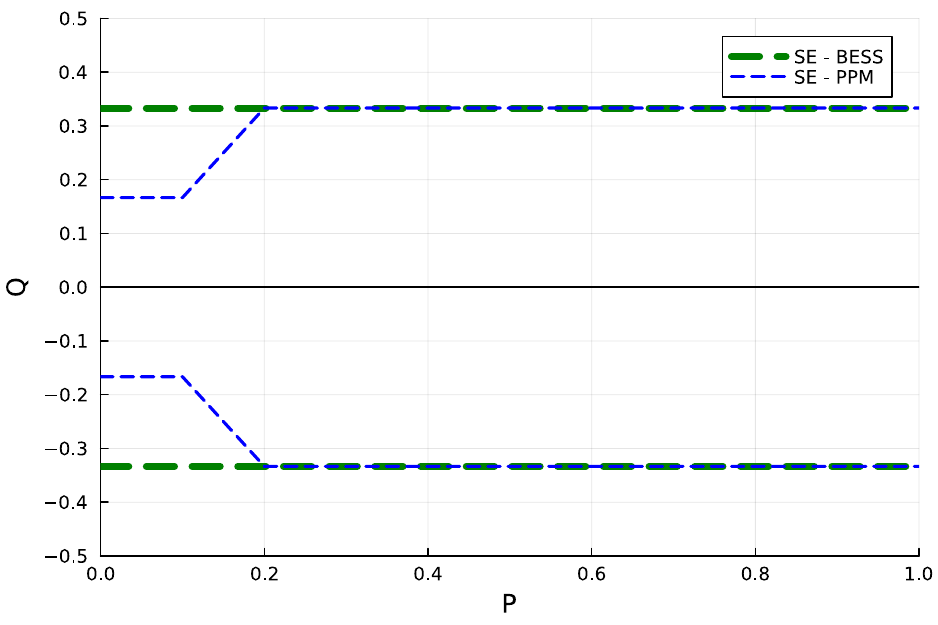}
	\caption{Reactive power requirements for PPMs and Battery Energy Storage Systems (BESS) according to the grid code suggestion made by Swedish TSO Svenska Kraftnät~\cite{SvK_utkast_2024}. BESS requirements apply for charging and discharging.}
	\label{fig:svk_utkast_ppm}
\end{figure}

The Swedish transmission system operator, Svenska Kraftnät, has suggested a new grid code~\cite{SvK_utkast_2024} to the Swedish electricity market regulator, which would amend the difference in reactive power requirements relative to the other Nordic grid codes. 
For SGs, the requirement would shift from $Q_\textrm{max}=1/3 P_\textrm{gen}$ and $Q_\textrm{min}=1/6P_\textrm{gen}$ to $Q_\textrm{max}=1/3P_\textrm{max}$ and $Q_\textrm{min}=1/6 P_\textrm{max}$. 
For PPMs, $Q_\textrm{max}$ and $Q_\textrm{min}$ would follow the profile shown in \cref{fig:svk_utkast_ppm} according to the suggestion. 
In addition, the Swedish grid code suggestion also specifies that, in the event of low or no active power PPM output, reactive power capabilities may not be reduced by disconnection or de-synchronization without a technically valid reason. 
This operation has also been referred to as STATCOM operation by the Norwegian grid code and is currently a capability requirement (instead of a standard operation requirement) for PPMs in Norway~\cite{Stattnet_2024}.

\subsubsection*{Grid code configurations}
Three grid code configurations are considered: %
\begin{itemize}
    \item The \emph{PPM in Q control} configuration represents current practice, whereby PPMs regulate the reactive power exchange at the Point Of Connection (POC) to zero during normal operation.
    \item In the \emph{PPM in V control} configuration, PPMs are in voltage control mode by default during normal operation. The reactive power capability is defined as shown in \cref{fig:grid_code}. The PPM voltage control point is at the LV side of the park transformer.
    \item The \emph{Updated grid code} configuration considers the new grid code suggested by Svenska Kraftnät. It increases the reactive power requirements for synchronous generators and PPMs (see \cref{fig:svk_utkast_ppm}), moves the control point of PPMs to the HV side of the park transformer, and requires STATCOM operation for PPMs.
    For the purposes of analyzing the effect of this grid code update, we apply the suggested Swedish grid code to the PPMs in all of the Nordic countries.
\end{itemize}

\subsubsection*{Synchronized generation} \label{sec:synchronized_generation}
The CE-ESM only models generation data per bus, timestep, and generation type; it does not model any the quantity of synchronized generation. 
As a consequence of this if, for example, there is power generation of 10~MW at a node, this could be produced by a generator with size $P_\textrm{max} = 10 ~\textrm{MW}$ running at full capacity or by a generator with size $P_\textrm{max}=20 ~\textrm{MW}$ running at half-load. 
As the reactive power requirements outlined in \cref{sec:reactive_power_req} depend heavily on the synchronized capacity of the generation unit, this has a significant impact on the available dynamic reactive power.

To account for this, we make the assumptions listed in \cref{tab:synchronized_generation}, in order to estimate the synchronized generation capacity at each timestep, bus, and for each generation type.
$P_\textrm{gen}$ is the actual generation output during the given timestep, $P_\textrm{max}$ is the annual peak generation, and $P_\textrm{en}$ is the synchronized generation capacity at the given timestep. 
The 80\% load factor for hydropower in the Nordic countries, which is based on \citet{Persson_2018}, allows for the provision of FCR and FRR. 
For PPMs, we assume that at $\geq$20\% generation the PPMs are fully synchronized, while for lower levels of generation we assume that individual generation units in the cluster of generation units have stopped producing, with a load factor of 50\% on the remaining units. 
For other (non-hydropower) synchronous generation and for HVDC systems, we assume that any generation at the bus and timestep means full synchronization of the buses' generation unit of that type. 

\begin{table}
    \caption{Overview of synchronized generation assumptions.}
    \label{tab:synchronized_generation}
    
    \begin{tabular*}{\tblwidth}{m{8em} m{15em}}\toprule
         Generation type& Synchronized generation assumption\\\midrule
        Hydropower& $P_\textrm{en} = P_\textrm{gen}/0.8$\\
         ~ \newline Wind \& solar& $P_\textrm{en} = \begin{cases}
			P_\textrm{max}, & \text{if $P_\textrm{gen}/P_\textrm{max} > 0.2$}\\
            2P_\textrm{gen}, & \text{otherwise}
		 \end{cases}$\\
 Other synchronous generation& $P_\textrm{en} = \begin{cases}
			P_\textrm{max}, & \text{if $P_\textrm{gen} > 0$}\\
            0, & \text{otherwise}
		 \end{cases}$\\
        ~ \newline HVDC& $P_\textrm{en} = \begin{cases}
			P_\textrm{max}, & \text{if $P_\textrm{gen} > 0$}\\
            0, & \text{otherwise}\\
		 \end{cases}$\\ \bottomrule
    \end{tabular*}
\end{table}

\subsection{Connection of generation to the transmission grid}
The reactive power requirements are generally set at the POC, unless agreed otherwise by the generation owner and TSO.
This POC often lies in the distribution grid, especially in the case of PPMs. 
However, we do not model the distribution grid. 
To consider the impedance between the POC and the transmission grid, we radially connect generation to the transmission grid with a transformer (trafo) and line impedance. Each generation type has its own radial connection at each bus.

\subsubsection*{Synchronous generators \& HVDC}
We assume that all synchronous generators (SGs) and HVDC stations are connected via a single trafo, as shown in \cref{fig:synch,fig:hvdc}.
The voltage set-point of the unit is chosen such that the reactive power injection into the transmission grid in the pre-contingency state is zero.

\subsubsection*{Power park modules}
\begin{figure}
\centering
\begin{subfigure}{\columnwidth}
\centering
\includegraphics[width=0.53\columnwidth]{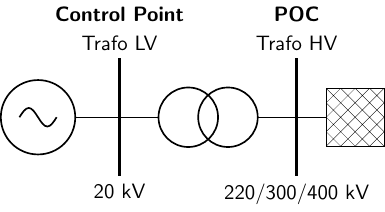}
\caption{Synchronous generator}
\label{fig:synch}
\end{subfigure}\hspace{10mm}

\begin{subfigure}{\columnwidth}
\centering
\includegraphics[width=0.51\columnwidth]{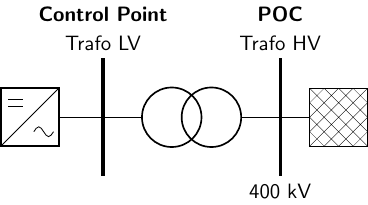}
\caption{HVDC}
\label{fig:hvdc}
\end{subfigure}\hspace{10mm}

\begin{subfigure}{\columnwidth}
\centering
\includegraphics[width=0.8\columnwidth]{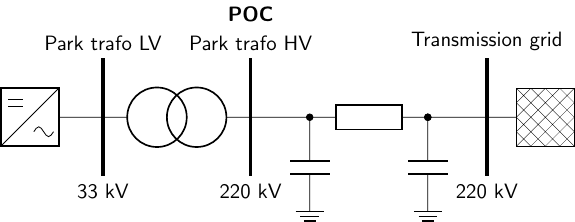}
\caption{220~kV Power park module}
\label{fig:ppm_220}
\end{subfigure}\hspace{10mm}

\begin{subfigure}{\columnwidth}
\centering
\includegraphics[width=\columnwidth]{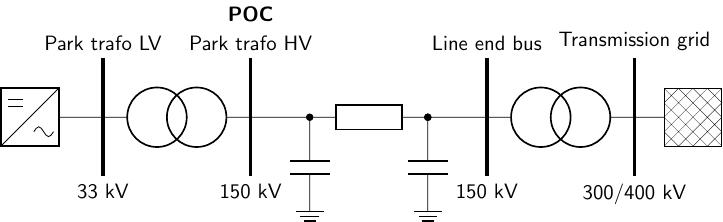}
\caption{300/400~kV Power park module}
\label{fig:ppm_300_400}
\end{subfigure}\hspace{10mm}

\caption{Grid connections by generation type and voltage level.\label{fig:grid_connections}}
\end{figure}

Power park modules are generally electrically more distant and connected to the distribution grid.
We apply two distribution grid structures to PPMs, depending on the voltage level of the transmission grid. 
For 220~kV, we use the structure shown in \cref{fig:ppm_220}, while for 300~kV and 400~kV we use the structure depicted in \cref{fig:ppm_300_400}.
We model the distribution grid line and transmission grid trafo as being perfectly shunt-compensated in the pre-contingency state. 
The voltage control point is either at the LV side of the park trafo (\emph{PPM in V control}) or at the HV side of the park trafo (\emph{Updated grid code}).
The voltage set-point at the voltage control point is chosen such that no reactive power is injected into the line at the POC. 
As the line and trafo are perfectly compensated, no reactive power injection at the POC means that there is no reactive power injection into the transmission grid.

\subsubsection*{Distribution grid parameters}
\Cref{tab:distribution_config} summarizes the line and trafo parameters used in \cref{fig:grid_connections}. 
The transmission line, park trafo, and grid trafo are scaled in accordance with the ratio of the maximum generation of the generation type at that bus and the distribution line capacity. 
A power factor of 0.9 is assumed for the purposes of scaling proportionally to the generator size. 

\begin{table*}[width=\textwidth]
  \caption{Configuration of distribution grid connections for generators.} \label{tab:distribution_config}
  \begin{tabular*}{\tblwidth}{@{} m{3.6cm}  m{1cm} m{1.48cm} *{5}{m{1cm}} m{3cm} @{}}\toprule
 Generation type & $X_\text{park trafo}$ [p.u.]&   Line length [km]& $L_\text{line}$ [mH/km]& $C_\text{line}$ [µF/km]& $R_\text{line}$ [$\Omega$/km]& $S_\text{line, nom}$ [MVA]&$X_\text{grid trafo}$ [p.u.]&Control point position\\\midrule
 Synchronous generator  & - &  - & - & - & - & - & 0.15  & Grid trafo LV\\
 Onshore PPM 220~kV &  0.15 & 20  & 1.56 & 0.01 & 0.049 & 440 & -  & Park trafo LV/ HV \\
 Onshore PPM 300/400~kV & 0.15 & 20   & 1.56 & 0.01 & 0.049  & 300 & 0.15 & Park trafo LV/ HV\\
 Offshore Wind 220~kV &  0.15 &  50 & 0.37 & 0.19&0.040  &475  & - & Park trafo LV/ HV\\
 Offshore Wind 300/400~kV & 0.15 & 50 & 0.37 & 0.19 &0.040 &475 & 0.15 & Park trafo LV/ HV\\ 
 HVDC & - & - & - & - & - & - & 0.15 & Grid trafo LV\\ \bottomrule
  \end{tabular*}
\end{table*}

%% file: GridModel.tex
\section{Case study: The Nordic synchronous power system} \label{sec:case_study}

\subsection{Grid dataset}
\begin{figure}
	\centering
		\includegraphics[width=0.9\linewidth]{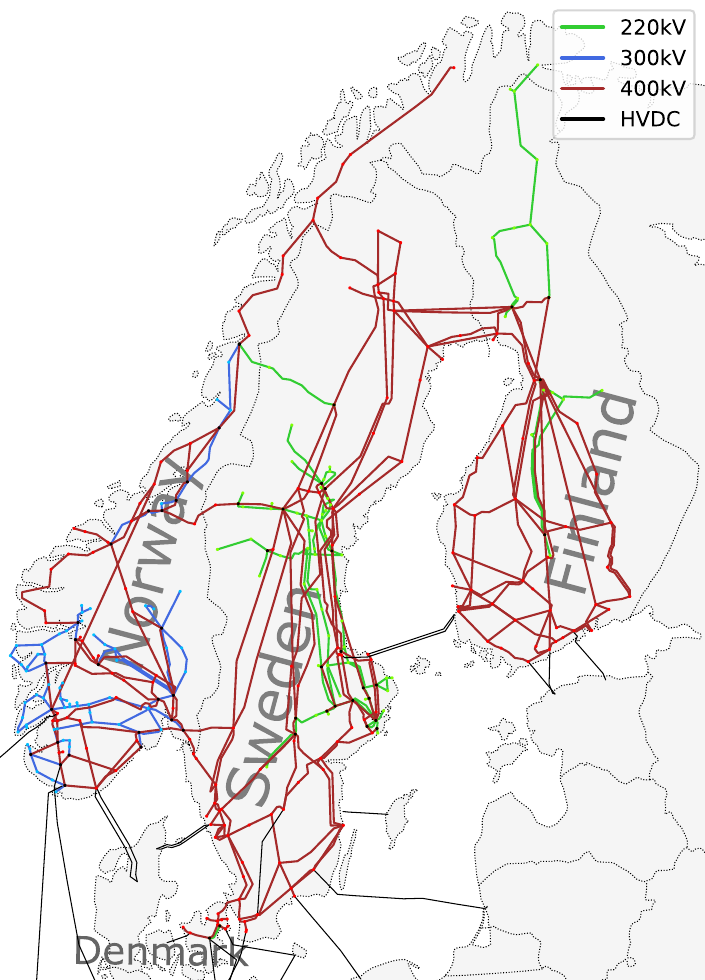}
	\caption{A map of the bus, HVAC, and HVDC line data used in this work.}
	\label{fig:grid_map_future}
\end{figure}

Our grid dataset is based on that of \citet{Hodel_2024a}, which uses data from the public ENTSO-E Grid map~\cite{ENTSOE_Map}.
For the future transmission system, we add and upgrade grid lines as outlined in the plans of the Nordic TSOs' long-term development plan and the Swedish TSOs' long-term development plan~\cite{NGDP_2025,Svk_lma_2024}.
A map of the lines and buses is presented in \cref{fig:grid_map_future}.
Further details and validation of the grid dataset can be found in \citet{Kuhrmann_2025}.

\subsection{Load and generation data} \label{sec:method_load_and_gen_data}

\begin{figure}
	\centering
		\includegraphics[width=\linewidth]{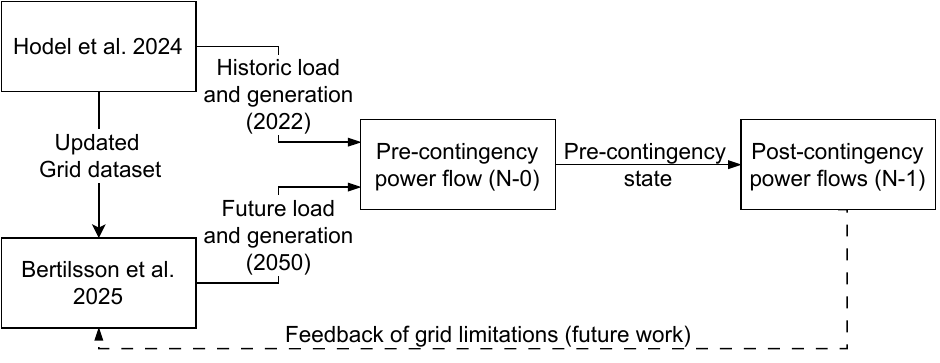}
	\caption{Flowchart of the load and generation data.}
	\label{fig:flowchart}
\end{figure}

Load and generation data are required for each bus and timestep for the contingency analysis.
\Cref{fig:flowchart} gives an overview of how the load and generation data for the historic and future scenario are sourced. 
For the historic scenario, the load and generation data are from \citet{Hodel_2024a}, which is based on published ENTSO-E data for Year 2022.
For the load and generation data for the future scenario, we use the results from the capacity expansion optimization model for Year 2050 presented by \citet{Bertilsson_2026}.
In both scenarios, the generation and load are connected to the nearest transmission grid bus.

\begin{figure*}[tp]
\centering
\begin{subfigure}[t]{0.4\linewidth}
    \centering
    \includegraphics[width=\linewidth]{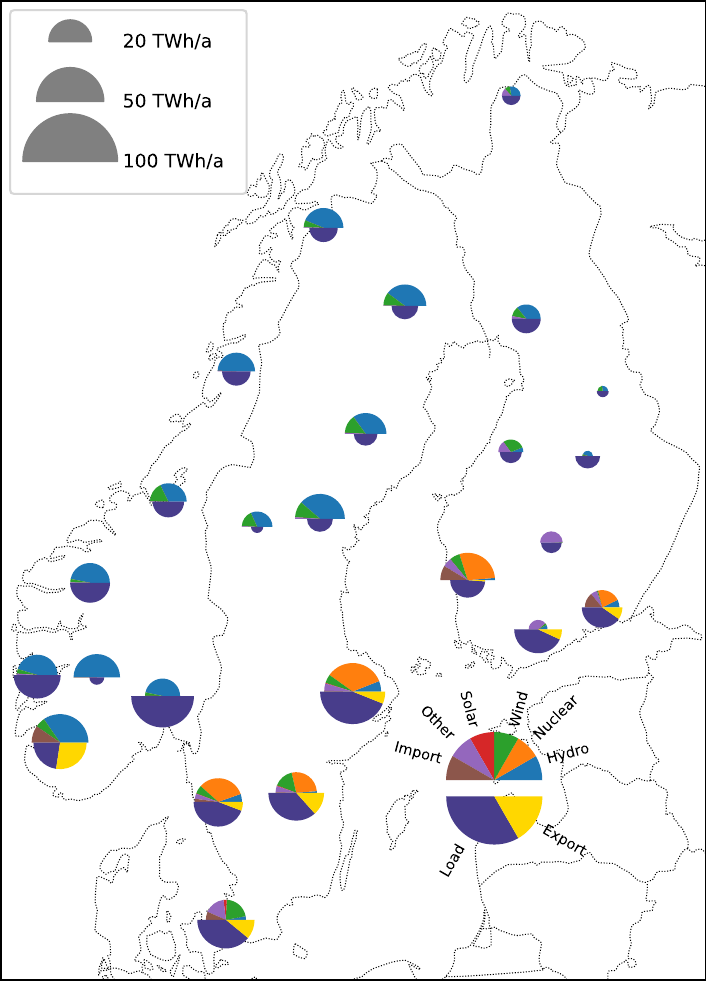}
    \caption{2022 - Historic scenario}
\end{subfigure}\hfil%
\begin{subfigure}[t]{0.4\linewidth}
    \centering
    \includegraphics[width=\linewidth]{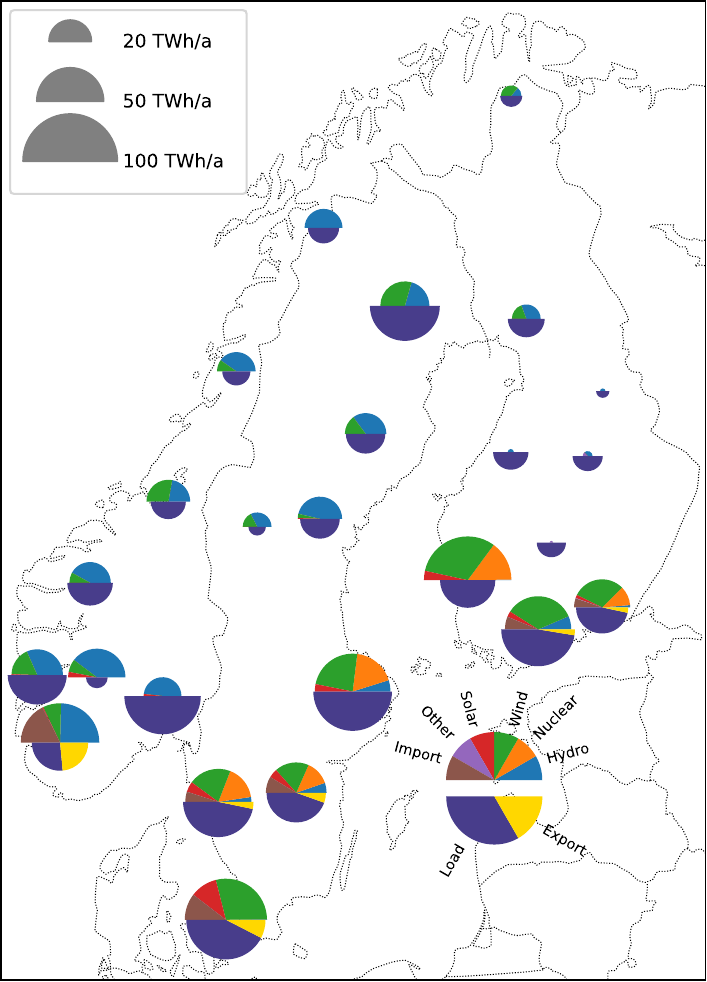}
    \caption[b]{2050 - Future scenario}
 \end{subfigure}
 \caption{Total annual load and generation levels in the historic and future scenarios. Buses are geographically clustered down to 25 nodes for visual clarity.}
 \label{fig:comparison_future-now}
\end{figure*}

The modeling carried out by \citet{Bertilsson_2026} assumes a significant increase in electricity consumption in the Nordic countries from around 350~TWh/a to around 650~TWh/a. 
An overview of the locations of load and generation both now and in the future scenario is provided in \cref{fig:comparison_future-now}.
The current system relies primarily on hydropower and nuclear power for electricity generation, while the future system sees a significant increase in electricity generation from wind power.

%% file: Results.tex
\section{Results} \label{sec:results}
\subsection{Comparison of grid code and extreme voltage automation configurations}\label{sec:comparison_GC}
The number of simulation runs that have voltage violations depends heavily on the grid code requirements on generators and on whether EXA is included. 
\Cref{tab:violations_overview} summarizes the different grid code and EXA configurations as well as the resulting numbers of simulation runs with voltage violations for both the historic and future scenarios.

\begin{table}
\centering

    \caption{Numbers and shares of voltage-insecure simulation runs for different grid code and EXA configurations in the future and historic scenarios.}
\begin{tabular}{@{} m{5.0em} >{\centering\arraybackslash}m{6.4em} *{2}{>{\centering\arraybackslash}m{5em}} @{}}
  \toprule
  EXA configuration & PPM in Q control & PPM in V control & Updated grid code \\
  
  \midrule
  & \multicolumn{3}{c}{Historic scenario} \\
  \cmidrule{2-4}
  No EXA & 134 (1.6\%) & 95 (1.2\%) & 74 (0.9\%) \\
  With EXA & 84 (1.0\%) & 62 (0.8\%) & 46 (0.6\%) \\

  \midrule
  & \multicolumn{3}{c}{Future scenario}\\
  \cmidrule{2-4}
  
  No EXA & 1548 (17.4\%) & 365 (4.1\%) & 236 (2.7\%) \\
  With EXA & 1282 (14.4\%) & 143 (1.6\%) & 95 (1.1\%) \\
  
  \bottomrule
\end{tabular}
\label{tab:violations_overview}
\end{table}

\subsubsection*{Historic vs. future scenarios}
\Cref{tab:violations_overview} shows that both the historic scenario and future scenario have voltage-insecure simulation runs in all of the grid code and EXA configurations.
As the actual system in Year 2022 was designed to be N-1 secure, which is a key operational criterion for the Nordic TSOs~\cite{Fingrid_2021}, the number of voltage-insecure simulation runs with the \emph{PPM in Q control} and \emph{With EXA} configuration should theoretically be zero.
However, some contingencies leading to voltage violations remain, due to several modeling limitations, including the exclusion of FACTS devices currently installed in the grid, inaccuracies related to the redistribution of generation and load to substations, the exclusion of sub-transmission networks, and the limited accuracy of the grid dataset.

When comparing the numbers of voltage-insecure simulation runs between the future and historic scenarios in \Cref{tab:violations_overview}, the future scenario has more voltage-insecure simulations than the historic scenario across all grid code and EXA configurations, indicating that it is less N-1 voltage-secure. 
This is the result of multiple changes that occurred between the two systems, including: an increased share of PPMs that are electrically further from the grid and, thus, provide weaker voltage support; generators in new locations; higher loading on the grid; changes in flow patterns; and more-dynamic operation of generation units.

\subsubsection*{Impacts of grid code variations}  
The results also highlight the importance of voltage control for PPMs. 
While the impact of PPM voltage control in the historic scenario is weaker, in the future scenario it is highly significant.
Such a difference between the scenarios is to be expected, as the share of inverter-connected generation increases significantly from the historic to the future scenario (see \cref{fig:comparison_future-now}).

A difference between the \emph{PPM in V control} and \emph{Updated grid code} configurations is evident in \cref{tab:violations_overview}.
The shares of voltage-insecure simulation runs in both the historic and future scenarios are reduced due to the additional dynamic reactive power provided by generation units, as well as the PPM control point being closer to the transmission grid.

\subsubsection*{Impact of EXA}    
Finally, the results show the impact of EXA on the number of voltage-insecure simulations. 
In both the future and the historic scenario, the inclusion of EXA exerts a significant effect on reducing the number of voltage-insecure simulations. 
In both scenarios, the static reactive power provided by shunt capacitors and reactors can make up for some of the dynamic reactive power deficiencies. 
However, some contingencies remain, where either insufficient shunt capacitors and reactors remain for EXA activation or where EXA activation is unable to return the voltage to its nominal range.

\subsection{Contingencies according to voltage-insecure frequency in the future scenario} \label{sec:most_violating_cont}
\begin{figure*}
\centering
    \includegraphics[width=\linewidth]{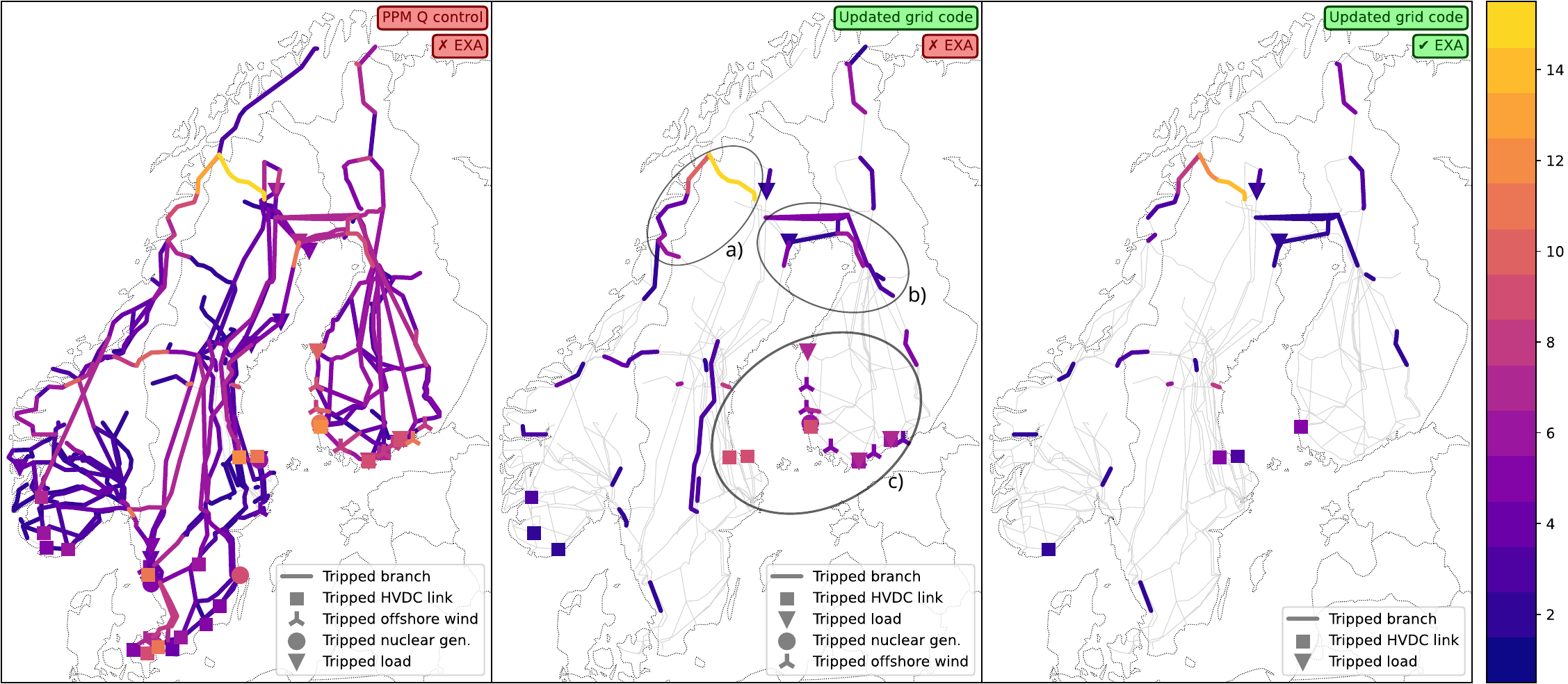}
 \caption{Numbers of flow states in which each contingency causes voltage violations in the future scenario. 
 The left panel shows the \emph{PPM in Q control} and \emph{No EXA} configuration. 
 The middle panel shows the \emph{Updated grid code} and \emph{No EXA} configuration. 
 The right panel shows the \emph{Updated grid code} and \emph{With EXA} configuration. 
 Contingency Clusters a), b), and c), are highlighted in the middle panel.}
 \label{fig:map_contingency_freq}
\end{figure*}

\Cref{fig:map_contingency_freq} provides an overview of the numbers of analyzed flow states in which each contingency causes violations for three different grid code and EXA configurations in the future scenario.

In \cref{fig:map_contingency_freq}, the contrast between \emph{PPM Q control} and \emph{Updated grid code}, both with \emph{No EXA}, is stark. 
In the \emph{PPM Q control} configuration, almost all contingencies are voltage-insecure in at least some flow states, while in \emph{Updated grid code} only a subset of contingencies is ever voltage-insecure. 
When comparing the \emph{Updated grid code} results in the \emph{With EXA} and \emph{No EXA} configurations, the difference is comparatively smaller.
Three clusters of contingencies that regularly lead to violations in the \emph{Updated grid code} and \emph{No EXA} configuration are highlighted and discussed in detail.

\subsection{Examples of contingencies without EXA}
The clusters of contingencies in the \emph{Updated grid code} and \emph{No EXA} configuration shown in \cref{fig:map_contingency_freq} are analyzed.

\subsubsection*{a) Northern Norway and Sweden}\label{sec:_a)}
One cluster of critical contingencies encompass the AC lines in northern Norway/ northern Sweden, which is marked as Cluster a) in \cref{fig:map_contingency_freq}. 
Voltage violations usually occur in these contingencies, as there are few alternative paths for electricity flow when they are tripped. 
As the sub-transmission grid is not modeled, there is only one very long alternative path, which raises contingencies that are not voltage-secure or do not converge. 
The line loads on this part of the grid may need to be reduced, as limited alternative paths for post-contingency flow are available, or sources of dynamic reactive power may need to be installed.

\subsubsection*{b) Northern Sweden and Finland}\label{sec:_b)}
\begin{figure}
	\centering
		\includegraphics[width=\linewidth]{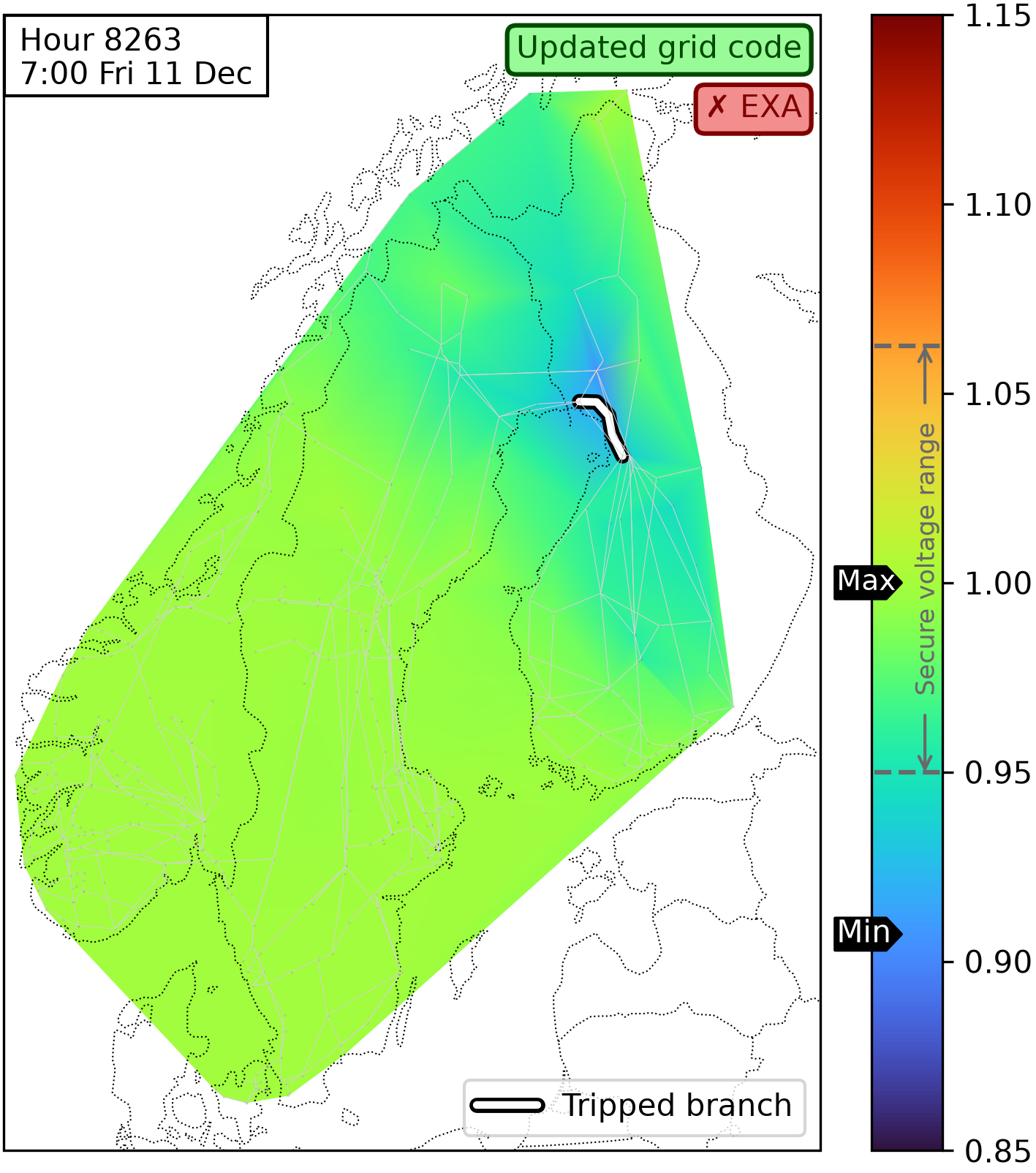}
	\caption{Example of post-contingency under-voltage caused by a line-tripping contingency under one critical flow state from Cluster b).}
	\label{fig:map_ex_b)_Voltages}
\end{figure}

\begin{figure}
	\centering
		\includegraphics[height=1.11\linewidth]{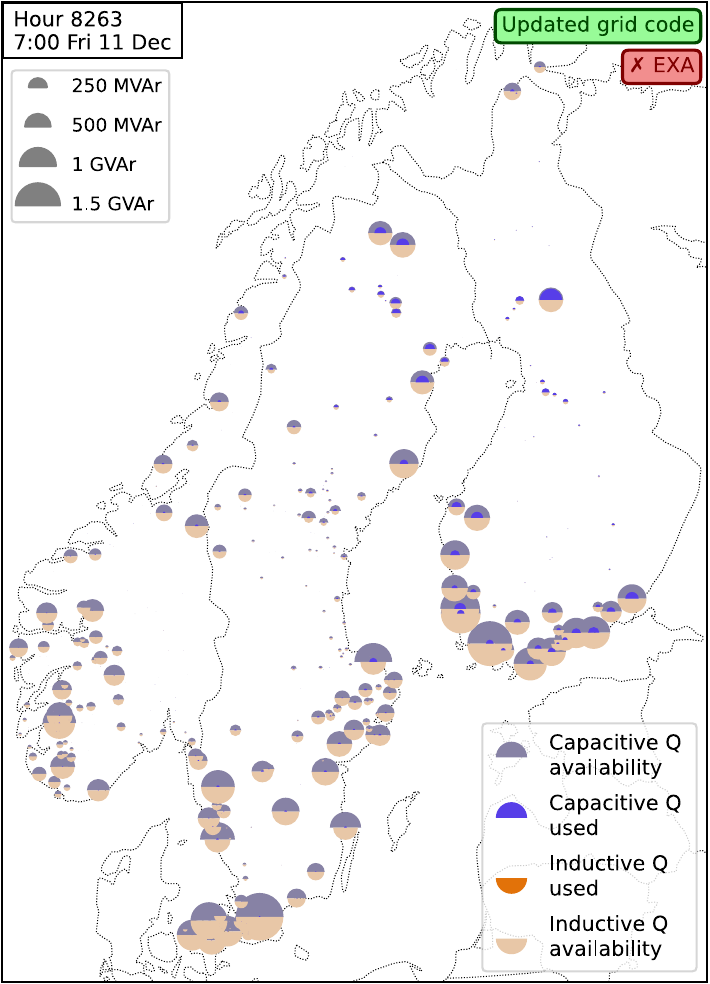}
	\caption{Map showing availability and usage of dynamic reactive power for an example contingency and flow state from Cluster b). 
    Buses are geographically clustered down to 300 nodes for visual clarity.
    }
	\label{fig:map_ex_b)_dyn_q}
\end{figure}

Another cluster of critical contingencies are the AC lines crossing between Sweden and Finland, marked as b) in \cref{fig:map_contingency_freq}. 
The post-contingency voltages for an example contingency from Cluster b), where a line trips during an example flow state, are shown in \cref{fig:map_ex_b)_Voltages}. 
The voltage in large parts of Finland drops significantly as the loading on the remaining lines between Sweden and Finland increases. 
The dynamic reactive power levels that are available and utilized during the contingency are shown in \cref{fig:map_ex_b)_dyn_q}. 
The dynamic reactive power in northern Finland is fully utilized but is insufficient to maintain the voltages within the voltage bounds. 
There is still dynamic reactive power available in southern Finland, although it is electrically too distant from the disturbance to be useful. 

To resolve this cluster of contingencies, operational or investment strategies, or both, can be considered. 
As an operational strategy, the pre-contingency flow in the AC lines crossing from Sweden to Finland could be reduced to reduce the size of the disturbance. 
For new investments, an EXA system, if not already in place, could be installed. 
Furthermore, the grid in northern Finland could be strengthened by installing additional dynamic reactive power sources, such as synchronous condensers or STATCOMs.

\subsubsection*{c) Southern Finland and HVDC to central Sweden} \label{sec:_c)}
\begin{figure}
	\centering
		\includegraphics[width=\linewidth]{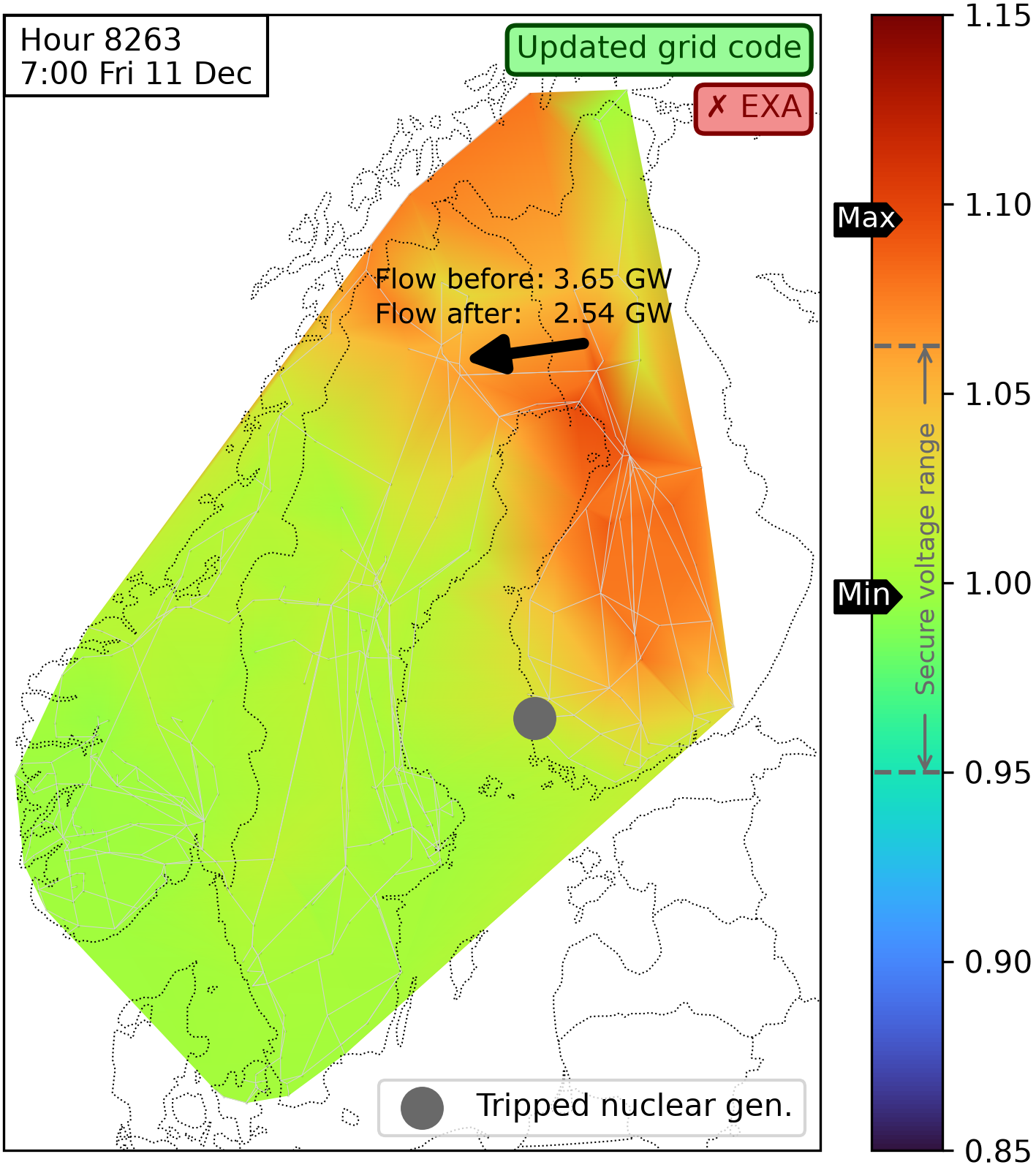}
	\caption{Example of post-contingency voltages for a contingency and flow state from Cluster c) caused by a generator tripping contingency. 
    The pre- and post-contingency flows across the Finnish-Swedish border are annotated.} 
	\label{fig:map_ex_c)_Voltages}%
\end{figure}

A second cluster of critical contingencies consists of large loads and generators in southern Finland, as well as the Fennoskan and Estlink HVDC lines, which cross to central Sweden and Estonia, respectively, marked as Cluster c) in \cref{fig:map_contingency_freq}.
The post-contingency voltages for an example contingency from this cluster where a nuclear generator trips during an example flow state are shown in \cref{fig:map_ex_c)_Voltages}. 
The contingency leads to a significant voltage increase in most of Finland. 
The voltage increases, as Finland previously exported active power to Sweden via their northern HVAC lines. 
Then, after the nuclear generator contingency in southern Finland, the hydropower-based FCR response resolves the active power imbalance. 
As this FCR hydropower response comes primarily from Norway and Sweden, Finland's exports drop by more than 1~GW, as annotated in \cref{fig:map_ex_c)_Voltages}. 
This reduction in active power flows leads to a severe voltage increase. 
As there are few reactive dynamic power resources in central and northern Finland, as is evident from the map of dynamic reactive power for the previous example contingency (\cref{fig:map_ex_b)_dyn_q}), it becomes difficult to regulate the voltage after the generator trips. 

Contingency Cluster c) has the same root cause as Cluster b), as both are voltage-insecure in the \emph{No EXA} configuration due to the lack of sufficient dynamic reactive power in northern Finland.

\subsection{Example contingencies with EXA}
When the EXA system is included, the contingencies that lead to voltage-insecure situations change. 
When comparing the contingencies and the frequencies with which they lead to voltage insecurity in \cref{fig:map_contingency_freq} in the \emph{Updated grid code} between the \emph{No EXA} and \emph{With EXA} configurations, it is clear that Cluster c) almost disappears and Cluster b) shows fewer flow states under which the contingencies are voltage-insecure.
Cluster a) largely remains in the \emph{With EXA} configuration, primarily because EXA does not change the situation that few alternative paths exist for the power to flow after the lines in the sparse grid around the cluster trip.

\begin{figure}
	\centering
		\includegraphics[width=\linewidth]{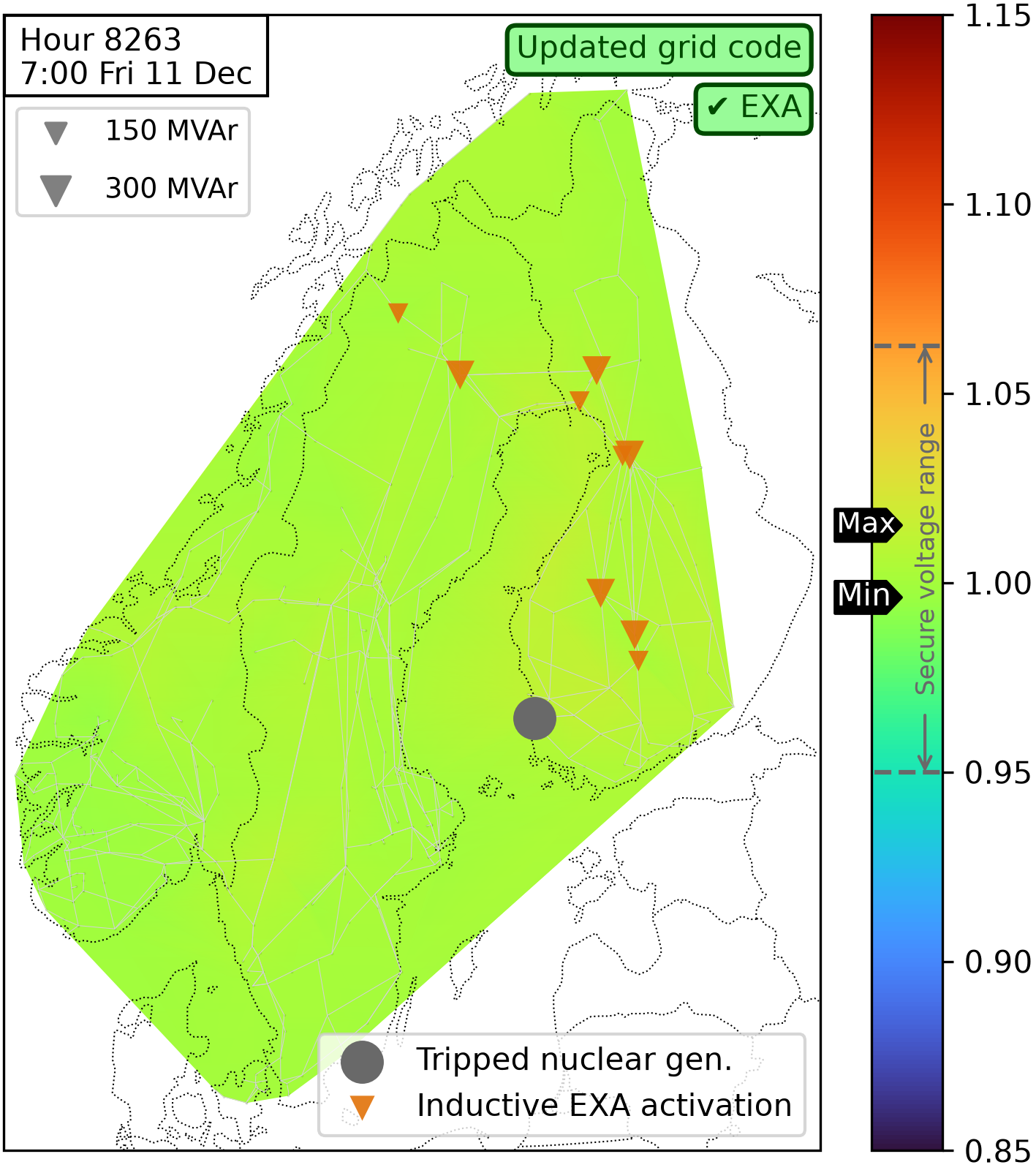}
	\caption{Example of post-contingency voltage containment due to EXA activation. The same contingency and flow state as in \cref{fig:map_ex_c)_Voltages} is shown.} 
	\label{fig:map_ex_c)_with_EXA}
\end{figure}

The example contingency from Cluster c) is shown in \cref{fig:map_ex_c)_with_EXA} in the \emph{With EXA} configuration.
The figure shows that with the EXA system, the voltage deviation in this contingency is contained effectively with no remaining over-voltage.
The EXA system activates at nine buses throughout Finland and northern Sweden for a total of 2,250~MVAr in connected inductive shunts.

This example highlights how the EXA system can stabilize contingencies that lead to large flow changes and, thus, to large voltage deviations.
It has a similar effect on line contingencies, although in general its effectiveness depends heavily on the remaining locally available shunt capacitors or reactors.

\subsection{Hydropower assumptions in the Capacity Expansion - Energy System Model} \label{sec:res_hydropower_assumptions}
The future scenario presented above was created by a CE-ESM with no operational minimum dispatch limits for hydropower. 
This means that the to the CE-ESM optimization uses the flexibility provided by hydropower to its limits to balance wind power variations.
However, hydropower operation comes with practical limitations.
Historically, hydropower has provided many frequency-stabilizing services by providing inertia, FCR, and FRR.
As such, certain minimum dispatch levels have been required for hydropower to be able to fulfill these needs.
However, with a changing ancillary service market, many of these services can also be provided by Battery Energy Storage Systems (BESS).
Another limitation associated with hydropower operation is environmental restrictions that influence operational freedom by restricting permitted dam levels and minimum or maximum flows.

To approximate these limitations in hydropower dispatch, a separate CE-ESM optimization is run in which a minimum generation level of 10\% is required for hydropower generation in each bus and each timestep.
10\% is chosen, as it is in line with historical hydropower production minimums per price area.
In \cref{sec:alt_hydropower_limits}, we present the results for this alternative future scenario with minimum hydropower dispatch. 
The results present both the numbers of voltage-insecure contingencies for different grid code and EXA configurations and a map of the contingencies showing in how many flow states they are voltage-insecure.

The results show that the number of voltage-insecure contingencies decreases significantly by about 25\%--65\% for all configurations compared with the future scenario presented in \cref{sec:results}.
This significant reduction highlights the sensitivity of the results to hydropower assumptions.
Nonetheless, the trends between configurations described in \cref{sec:comparison_GC} still hold.
Furthermore, while the clusters indicated in \cref{fig:map_contingency_freq} have significantly weakened, they are still present.

%% file: Discussion.tex
\section{Discussion} \label{sec:discussion}
In this section, we first discuss the modeling limitations (\cref{sec:disc_limitations}), followed by the implications for grid operation and CE-ESM optimization (\cref{sec:disc_implications}).

\subsection{Modeling limitations} \label{sec:disc_limitations}
\subsubsection*{N-1 assumptions}
A core assumption in this study is that the pre-contingency voltages are 1~p.u., which is not accurate in reality, as the voltages are constantly changing. 
This assumption is necessary because no data on the placement, sizing and operation of shunt capacitors and shunt reactors in the Nordic Countries are publicly available.  
Furthermore, no information is available on how future shunt capacitors and reactors would be placed.
As our focus is primarily on the effects of contingencies, we believe the 1~p.u. assumption to be useful.

The transmission grid model only includes lines with voltages in the range of 220--400~kV, thereby excluding all sub-transmission lines. 
This could significantly change the impacts of contingencies, as post-contingency flow paths would behave differently if sub-transmission lines were available. 
This effect might reduce the impacts of contingencies, as there are more alternative paths for the power to flow.

We assume a fully operational transmission grid, with no lines out of operation for maintenance. 
This is an over-estimation of the actual number of available lines, which makes the results too optimistic. 
Lines can be out of operation for planned reasons, such as maintenance, or for unplanned reasons such as outages or faults. 
While maintenance is usually scheduled for periods with low expected flows, longer unplanned outages can happen at any time, with the system required to return to N\nobreakdash-1 stability in order to be ready for further contingencies.

\subsubsection*{Flow state and contingency selection}
Due to computational limitations, not all flow states are evaluated with respect to N\nobreakdash-1 security. 
Instead, only 18 special flow states are evaluated. 
As N\nobreakdash-1 insecurity can be a local problem, and flow states show considerable local differences, some voltage-insecure contingencies that are dependent upon specific local conditions may not be found.

While the considered contingencies are broad, some assumptions are made.
Double circuit lines are assumed to only lose one circuit, even if both circuits are located on the same towers.
The only load contingencies that are included involve future hydrogen electrolyzers. 
Currently, existing industrial loads (e.g., aluminum smelters) are considered in the FCR dimensioning process~\cite{Entsoe_FCR_2024} and they could therefore have significance for N\nobreakdash-1 stability. 
However, as no data on load size or location are available, we have included only future electrolyzers as load contingencies.
Only nuclear and offshore wind power generation contingencies are considered, as all other generation contingencies are assumed to be of smaller magnitude.
However, depending on the location, smaller contingencies could still have stronger impacts.

\subsubsection*{Static reactive power modeling}

Shunt capacitors and reactors are placed at each bus and sized to exactly produce voltages of 1~p.u. in the pre-contingency state. 
In addition, no reactive power is assumed to be exchanged between the distribution grid and the transmission grid in the pre-disturbance state. 
This is in agreement with the static reactive power exchange requirements between TSOs and DSOs~\cite{Svk_2024b}.

EXA activation is modeled to act with perfect centralized control: shunt capacitor and reactor activation is distributed well between all shunt units, with little risk of ``over-activation''. 
In reality, EXA activation is only locally controlled: once voltages go below a chosen threshold for a set amount of time, connection/disconnection is initiated. 
This risks ``over-activation'', as nearby units might connect/disconnect excessive reactive compensation, which might change, for example, an under-voltage situation to an over-voltage situation.
However, centralized control is likely not desirable in reality: desired EXA reaction times are at most a few seconds, centralized control would require extremely rapid state estimation to ensure adequate control decisions, and centralized control would reduce the robustness of the system, as it currently requires only local voltage measurements.
Therefore, we are likely over-estimating the true capabilities of EXA to some degree. 

When considering EXA activation, only shunt capacitor and reactor sizes of 150~MVAr are considered. This is limiting in cases where the grid is electrically very weak. 
Smaller sizes could be used in such places, which would allow for more fine-grained control of the voltage by the EXA reaction. 
Providing dynamic reactive power at such places is another option.

\subsubsection*{Dynamic reactive power modeling}
We make the following assumptions regarding how we model available dynamic reactive power:
\begin{itemize}
 \item Generation units supply as much reactive power as the grid code requires, rather than their outputs being based on what their technical limitations are.
 \item Generation units are assumed to be of type C or D (>10~MW).
 \item All dynamic reactive power is assumed to be available for contingencies.
 \item Synchronized generation is only assumed based on current generation and annual maximum generation levels.
\end{itemize}
Due to all of these factors, we could be both under- and over-estimating the dynamic reactive power available for handling a contingency event.
Nevertheless, we expect the broader conclusions to hold, as they are less sensitive to the exact quantity of available dynamic reactive power and much more sensitive to the location at which the dynamic reactive power is provided.

\subsubsection*{Voltage stability assessment}
We do not evaluate the remaining voltage stability margin. 
Even though Newton-Raphson power flow converges, the system is not guaranteed to be N-1 secure if the new operating point is very close to the nose (voltage-collapse) point of the PV curve. 
This is especially critical because EXA activation increases the degree of reactive compensation, leading to smaller voltage stability margins.
Future work should evaluate the remaining stability margins to provide a more reliable assessment of stability.

\subsubsection*{FCR allocation}
We assume that FCR is provided exclusively by hydropower, with all hydropower dispatch being evenly and proportionally scaled. 
This leads to larger line loading changes at the Sweden-Finland border after a load or generation disturbance occurs in Finland, as the vast majority of hydropower generation, and thus the vast majority of FCR in our model, comes from Norway and Sweden. 
We model Fingrid as providing 6.5\% of the Nordics' FCR on average, whereas in reality Fingrid needs to procure 20\% of the Nordics' FCR, of which at most one-third may come from trade with other TSOs~\cite{Entsoe_FCR_2024}. 
Still, as the majority of FCR is non-Finnish, we are likely not over-estimating the cross-border FCR flow after a generation or load contingency in northern Finland by much.

The importance of BESS for FCR is growing, with around 30\% of Swedish pre-qualified FCR capacity being BESS~\cite{Svk_FCR_2026}. Future work should investigate the impact of FCR provided by BESS on N-1 voltage security, especially in relation to BESS locations.

\subsubsection*{Historic data limitations}
The accuracy of the historic data reconstruction is limited, as the generation and load data were redistributed from bidding zone level to individual buses (for details, see \citet{Hodel_2024a}). 
In addition, assumptions as to conductor types were made, as the exact data are not publicly available (for details, see \citet{Kuhrmann_2025}).

\subsubsection*{Future data limitations}
The CE-ESM for Year 2050 described by \citet{Bertilsson_2026} has limitations as a data source, as future demand is difficult to predict. 
In addition, as the model of \citet{Bertilsson_2026} is a greenfield model in terms of wind and solar installations, it does not consider the currently installed wind parks in, for example, northern Finland. 
Furthermore, the 3-hour time resolution used leads to smoothed-out flows with less-extreme operating points, while the real loading conditions could be more severe. 

Comparisons between the historic and future scenarios should, therefore, be made with caution, although the broader trends and comparisons within each scenario should hold.

\subsubsection*{Hydropower modeling}
Although the future operational strategies and limitations for hydropower may change, the impact of a minimum hydropower dispatch assumption of 10\% for all time steps is tested. 
When aggregated to the entire system, this provides a closer match to the current minimum hydropower dispatch in the whole Nordic system.
One proxy for the realism of hydropower dispatch is the minimum hydropower generation level in the Nordic countries.
The minimum hydropower production level in the Nordic countries in the historic scenario (Year 2022) is 8.7~GW, in the future scenario it is 1.4~GW, while in this alternate future scenario with minimum hydropower dispatch it is 4.7~GW. 
However, it remains unclear as to how much of the historic minimum dispatch was influenced by energy market needs, frequency stability needs, and environmental limitations. 
This makes it difficult to say how low a realistic future minimum dispatch can be.

The results show a significant reduction in the number of voltage-insecure contingencies if a minimum hydropower dispatch level of 10\% is required at all times.
However, a minimum hydropower dispatch of 10\% significantly changes the amount of available dynamic reactive power, as all buses with hydropower generation then always have some level of available dynamic reactive power. 
This does not reflect reality, in that individual hydropower plants do have periods in which they do not produce any power and, thus, provide no dynamic reactive power.
Therefore, we choose to use primarily the optimization results without minimum hydropower production limitations, as this is the more conservative option from the perspective of dynamic reactive power availability.

Overall, the future operational strategies and limitations of hydropower remain unclear, which impacts the availability of dynamic reactive power especially at locations where other generation resources are limited. 
However, our conclusions on the impact of grid codes regarding reactive power and voltage control requirements, as well as the impact of EXA still hold for both hydropower modeling approaches.
Further work on modeling hydropower dispatch limitations in high-resolution CE-ESMs is needed.

\subsection{Implications} \label{sec:disc_implications}

\subsubsection*{Grid code implications}
The comparison between grid codes described in \cref{sec:comparison_GC} highlights the importance of the control mode of PPMs. 
If PPMs are in reactive power control mode or in unity power factor control mode they provide no voltage support in the case of a contingency. 
In other words, as the grid voltage drops, the internal voltage of the PPM also decreases to maintain a given setpoint of reactive power flow exchange. 
This significantly impacts the voltage stability of the grid, as shown by the large increase in the simulations that violate the voltage bounds in \cref{tab:violations_overview}, when comparing the historic and future scenarios.
It should be noted that the Swedish grid code currently requires an automatic changeover to voltage control mode if the control point voltage drops below 0.95~p.u. However, such a changeover to voltage control model is not required in the case of over-voltages. 
In addition, it could be the case that even though the voltage at the control point is $\geq$0.95~p.u., the voltage at the transmission grid falls below 0.95~p.u.
The importance of voltage control mode operation has also been highlighted in the report on the Year 2025 blackout in Spain and Portugal, as the Iberian system did not require voltage control mode for PPMs \cite{entsoe_2026}.

The results presented in \cref{sec:comparison_GC} also show that updating the grid code as suggested by Svenska Kraftnät to require that PPMs operate in STATCOM operation as well as to have a control point closer to the transmission grid helps to reduce the number of voltage violations.
Note that STATCOM operation requires inverters to always be synchronized and switching, which increases active power losses. 
These losses would lead to an increase in the price of electricity for electricity consumers. 
How this cost for STATCOM operation compares to other approaches to ensure voltage stability, such as TSO investments in FACTS devices, is not currently known and this should be investigated in future work.

\subsubsection*{Extreme voltage automation implications} %
The results show a significant reduction in the number of voltage-insecure simulations when EXA is included in both scenarios, with a stronger impact observed in the future scenario.
This importance of automated shunt capacitor and reactor connection/disconnection has also been highlighted by the ICS Investigation Expert Panel, which investigated the Year 2025 blackout in Spain and Portugal, as the Iberian system does not have shunt automation \cite{entsoe_2026}.
It is unclear as to whether the Nordic countries, apart from Sweden, have EXA but if they do not, they should consider adding it to their system integrity protection schemes (SIPS). 

An important system design decision is whether EXA should be relied on for N-1 voltage security or whether the system must remain N-1 voltage-secure independent of EXA's availability, with EXA serving as a backup system.
This decision determines how much additional dynamic reactive power source is needed to support the system.

We have also investigated the impact of EXA when not constrained by remaining shunt capacitors and reactors. 
In the future scenario with the \emph{Updated grid code}, this reduces the percentage of voltage-insecure simulations further from 1.1\% in the \emph{With EXA} configuration to 0.4\% when EXA is not constrained by the remaining capacities.
In the historic scenario with the \emph{PPM in V control} configuration, a reduction from 0.8\% to 0.2\% is found.
However, despite these apparent large increases in N-1 voltage security, these cases may not necessarily have sufficient voltage stability margins, and their stability levels need to be evaluated in future work.

In the \emph{With EXA} configuration, as well as when EXA is unconstrained by remaining capacities, we identify cases in which the grid is too electrically weak for 150~MVar-sized shunts, i.e., an under-voltage flips to an over-voltage if a 150~MVAr-sized shunt capacitor is connected. 
This primarily occurs in the more remote 220~kV parts of the grid.
In such weak grid locations, either smaller-sized EXA shunts or dynamic reactive power may be needed.

\subsubsection*{Contingencies} %
It is worth noting that voltage violations can occur far from the site of the contingency, as shown in the example in \cref{fig:map_ex_c)_Voltages}. 
Both under- and over-voltages can occur remotely. 
This phenomenon is primarily attributed due to a change in net generation (i.e., after a loss of generation, HVDC, or a loss of load), as the FCR that resolves the imbalance can be located far away from the contingency, which changes the loading in the grid that delivers the FCR to the location of the power imbalance. 
This finding may defy expectations that the impacts of contingencies on voltage stability are purely local.

\subsubsection*{Capacity Expansion - Energy System Model implications} \label{sec:disc_opt_model_implications}
As voltage violations occur even with an updated grid code and EXA, the CE-ESM optimization results are not N\nobreakdash-1 voltage-secure as is. 
In the absence of EXA, northern Finland in particular is affected, as insufficient dynamic reactive power leads to voltage violations (see the example in \cref{fig:map_ex_b)_Voltages}).
When including EXA, the number of voltage-insecure simulations is still not zero.
Especially in more remote and weaker parts of the grid, voltage-insecure situations remain: additional dynamic reactive power is likely needed to ensure that these contingencies are voltage-secure.
Different strategies can supply this dynamic reactive power, e.g., investments in sources of dynamic reactive power, such as STATCOMs. 
Alternatively, altering the locations of generation units may resolve voltage violations, as this measure also moves the dynamic reactive power that the generator units provide.
While considering these limitations when modeling electricity systems will likely not fundamentally change the optimal generation composition, it is still worth evaluating voltage stability aspects, to ensure the operability of a future electricity system.
This suggestion is in line with IEA Wind TCP Task 25, which recommends evaluating transmission network stability when conducting wind and solar integration studies~\cite{Goeransson_2026}.

%% file: Conclusion.tex
\section{Conclusion}\label{sec:conclusion}

In this work, we apply N-1 voltage stability analysis to a number of challenging flow states obtained from a CE-ESM that represents the Nordic transmission grid in Year 2050 with around 400 buses. 
We apply the same N-1 voltage stability analysis to a historic scenario for Year 2022 based on ENTSO-e data.
We model simplified collection grids for PPMs and consider different grid code requirements for the dynamic reactive power supplied by generators.
We assume unity pre-contingency voltages, and the modeled cases include those with and without EXA for shunt capacitors and reactors.
We apply HVAC, HVDC, generation, and load contingencies to several challenging flow states and evaluate the post-contingency voltages.

The results reveal the existence of N-1 voltage violations in the future scenario generated by the CE-ESM.
The number of voltage-insecure contingencies varies significantly depending on the grid code and EXA configuration.
The \emph{PPM in Q control} and \emph{No EXA} configuration leads to the highest number of voltage-insecure contingencies, with 17.4\% of the contingency and flow state combinations being voltage-insecure.
In contrast, the \emph{Updated grid code} and \emph{With EXA} configuration leads to the fewest voltage-insecure contingencies, at just 1.1\%.
These voltage insecurities need to be addressed to ensure that a future system is N-1 secure.

The results highlight that contingencies that affect areas with limited to no dynamic reactive power are voltage-insecure in many flow states. 
In the future scenario a heavily affected area is northern Finland, as the CE-ESM places few new generation units and, thus, little dynamic reactive power in this area.
When including the EXA response, however, some of these contingencies are resolved.

The results also show the impacts of different grid code requirements on the supply of dynamic reactive power from PPMs. 
They highlight the critical importance of the PPM control mode in the future scenario, whereby the voltage control mode leads to 1.6\%--4.1\% of simulations being voltage-insecure compared to 14.5\%--17.4\% when operating in reactive power control mode, depending on the EXA configuration. 
Such a substantial change is not observed in the historic scenario, as fewer PPMs are present in the current system.
In addition, the results emphasize the positive impacts of the \emph{Updated grid code} with PPM STATCOM operation requirements and PPM voltage control point position changes. 
These changes lead to further reductions in voltage-insecure simulations from 1.6\%--4.1\% to  1.1\%--2.7\%.
While requiring STATCOM operation does lead to fewer voltage violations, the additional active power losses and their associated costs should be evaluated.

EXA is shown to yield a very significant reduction in the number of voltage-insecure simulations.
While it is unclear if the non-Swedish parts of the Nordic system have EXA, if absent, TSOs should consider installing EXA due to its apprently strong impact.

Further studies on the remaining voltage stability margins and time-domain simulations of voltage deviations are required to fully understand the voltage security and stability levels of future scenarios.

%% file: Appendix.tex
\appendix
\section{Appendix}

\counterwithin*{figure}{part}
\counterwithin*{table}{part}
\stepcounter{part}
\renewcommand{\thefigure}{A.\arabic{figure}}
\renewcommand{\thetable}{A.\arabic{table}}

\subsection{Alternative hydropower operational limits} \label{sec:alt_hydropower_limits}

\Cref{tab:violations_future_hydro_min_disp} shows the number of voltage-insecure contingencies for the considered grid codes and EXA configurations with 10\% minimum hydropower dispatch.

\begin{table}
\centering
    \caption{Numbers and shares of voltage-insecure simulation runs for the different grid codes and EXA configurations in the future scenario with 10\% minimum hydropower dispatch.}
\begin{tabular}{@{} m{5.0em} >{\centering\arraybackslash}m{6.4em} *{2}{>{\centering\arraybackslash}m{5em}} @{}}
  \toprule
  EXA configuration & PPM in Q control & PPM in V control & Updated grid code \\
  
  \midrule
  & \multicolumn{3}{c}{Future scenario, min hydropower dispatch}\\
  \cmidrule{2-4}
  
  No EXA & 1,184 (13.3\%) & 174 (2.0\%) & 105 (1.2\%) \\
  With EXA & 861 (9.7\%) & 56 (0.6\%) & 38 (0.4\%) \\
  
  \bottomrule
\end{tabular}
\label{tab:violations_future_hydro_min_disp}
\end{table}

\begin{figure*}
\centering
    \includegraphics[width=\linewidth]{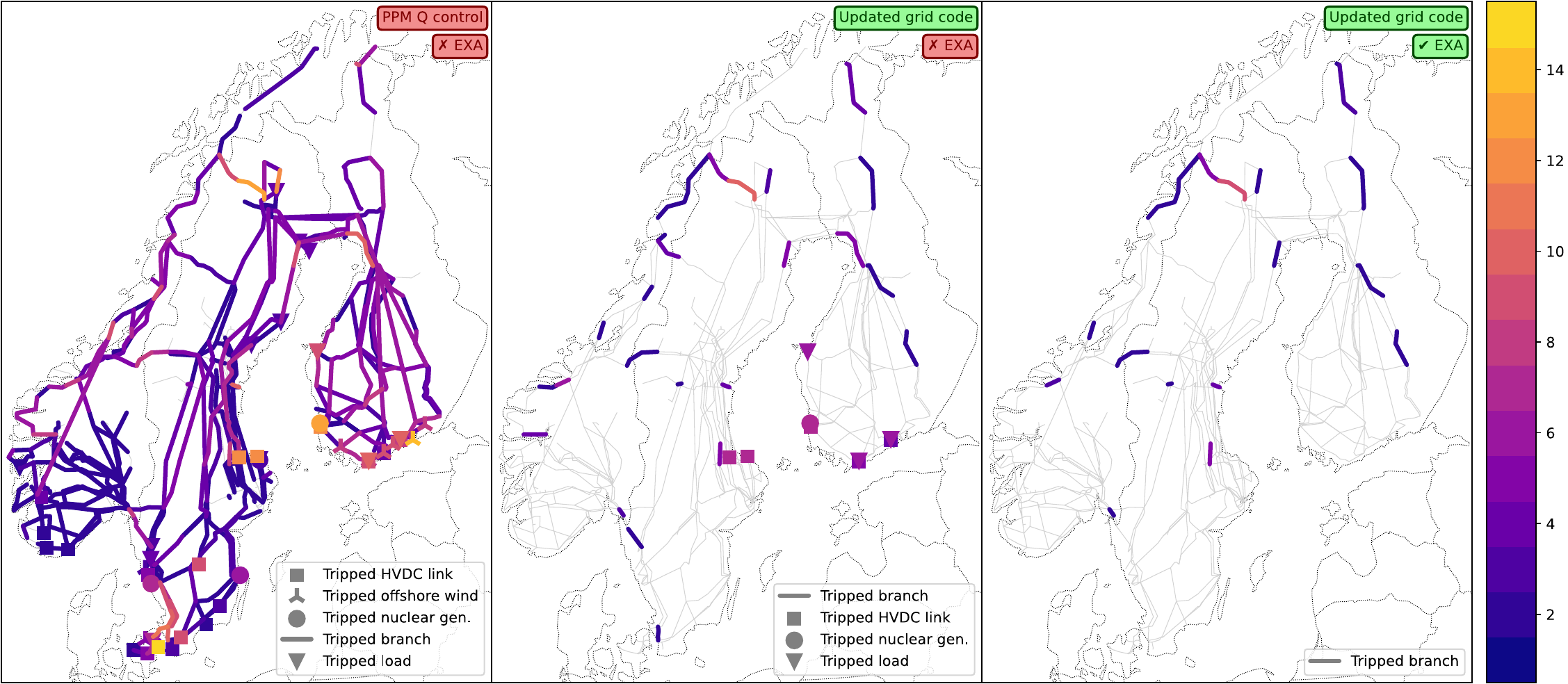}
 \caption{Number of flow states in which each contingency causes voltage violations in an alternate future scenario created with a 10\% minimum hydropower production constraint. }
 \label{fig:map_contingency_freq_min_hydro}
\end{figure*}

\Cref{fig:map_contingency_freq_min_hydro} presents the number of flow states in which each contingency causes voltage violations with the 10\% minimum hydropower dispatch (see \cref{fig:map_contingency_freq} for comparison).

